\documentstyle[aps,prl,epsf,floats]{revtex}

\def\Im{{\rm Im}}
\def\osc{{\rm osc}}
\def\sc{{\rm sc}}

\newcommand{\Postscript}[7]{
\begin{figure}[htb]
\begin{center} 
\vspace{#2truecm}
\begin{picture}(#1,#1)
\epsfxsize=#1truecm \epsfbox{#3}
\end{picture}
\parbox{#5truecm}{
\vspace{#4truecm}
\caption[figure1]{\small \it #6 }
\label{#7}}
\end{center}
\end{figure}
}

\begin{document}
\bibliographystyle{unsrt}
\newcommand{\be}{\begin{equation}}
\newcommand{\ee}{\end{equation}}
\newcommand{\bea}{\begin{eqnarray}}
\newcommand{\eea}{\end{eqnarray}}

\draft
\preprint{}
\twocolumn[\hsize\textwidth\columnwidth\hsize\csname@twocolumnfalse%
\endcsname

\title
{Complex Periodic Orbits and Tunnelling in Chaotic Potentials}

\author{Stephen C. Creagh and Niall D. Whelan}
\address{Division de Physique Th\'{e}orique\cite{cnrsbullshit}, 
IPN, 91406 Orsay Cedex, France.}

\date{\today}

\maketitle
\begin{abstract}
We derive a trace formula for the splitting-weighted density of states
suitable for chaotic potentials with isolated symmetric wells.
This formula is based on complex orbits which tunnel through classically
forbidden barriers. The theory is applicable whenever the tunnelling is
dominated by isolated orbits, a situation which applies to chaotic
systems but also to certain near-integrable ones. 
It is used to analyse a specific two-dimensional potential with chaotic 
dynamics. Mean behaviour of the splittings is predicted by
an orbit with imaginary action. Oscillations around this mean are obtained 
from a collection of related orbits whose actions have nonzero
real part. 
\end{abstract}

]

In this letter we develop a method for computing tunnelling effects in 
quantum states associated with chaotic regions of phase space.
While the influence of chaotic dynamics has been actively studied in the 
context of chaos-assisted tunnelling between EBK-quantised tori
\cite{cat,more}, much less  
attention has been paid to the problem of tunnelling between chaotic states. 
This is presumably because there is at present no semiclassical theory for 
individual chaotic eigenstates. However, there does exist a
well-developed theory for the spectral properties of such systems in terms 
of periodic orbits \cite{potissue}. We will show how this can be
extended to a calculation of spectral tunnelling averages using
complex periodic orbits.

Complex multi-dimensional trajectories were introduced in \cite{bbwbb} to
understand barrier penetration problems as motivated by quantum field 
theories. Complex periodic orbits were first used by Miller \cite{miller} 
in one-dimensional potentials, to derive splittings and resonance widths
and these calculations were extended in \cite{SF6} to the 
rotational spectrum of $\mbox{SF}_6$. In the context of chaotic 
maps, complex periodic orbits were used to calculate band gaps 
\cite{thebeef}. Complex trajectories in chaotic
maps have also been explored in the time domain \cite{shudo}.

A commonly used probe for tunnelling effects is to study the spectra
of two symmetric wells which have a barrier between them. When classical 
trajectories are localised within one well or the other, 
one finds that energy levels come in symmetric-antisymmetric pairs 
$E_n^{\pm}$, with small splittings between them. Denote
the splittings by $\Delta E_n$ and the mean levels by $E_n$, so that
$E_n^\pm=E_n\mp\Delta E_n/2$. A standard periodic orbit calculation
using real orbits yields a set of doubly-degenerate levels $E_n^\sc$,
approximating the mean levels $E_n$. The degeneracy arises because
each periodic orbit has a symmetric partner in the opposing well.
Therefore, the most na\"{\i}ve use of periodic orbit theory fails to
predict the existence of splittings. However, since 
including {\it complex} orbits does yield splittings in one dimension
\cite{miller,SF6}, we are motivated to pursue this approach in higher 
dimensions.

A literal extension of this work to chaotic problems is too
difficult because it involves finding tiny differences between
poles in traces or zeros in zeta functions.
Instead, the splitting-weighted density of states,
\begin{equation}\label{def} 
    f(E) = \sum_n \Delta E_n \; \delta(E-E_n).
\end{equation}
will provide an effective vehicle for evaluating splittings
directly from complex orbits. Analysis of $f(E)$ in terms of periodic 
orbits follows from approximating it by the difference,
\begin{equation}\label{diffN} 
    f(E) \approx N_{+} (E) - N_{-} (E),
\end{equation}
%
between the staircase functions $N_\pm(E)$ for even and odd states.
The approximation is valid if we use a resolution in
energy that is much greater than $\Delta E_n$. We next note that
$N_\pm(E) = -(1/\pi) \Im \int^E dE' \, g_\pm(E')$, where $g_\pm$ are
traces of the symmetry-projected Greens functions  \cite{robbins}
\be \label{irreps} g_\pm(E) = 
\sum_n{1\over E-E^\pm_n} \approx {1\over 2}
\sum_\gamma\chi_\pm(g_\gamma)A_\gamma e^{iS_\gamma/\hbar}.
\ee
The traces receive contributions from orbits
$\gamma$ which either close simply in phase space ($g_\gamma=I$) or 
close after applying a reflection operation ($g_\gamma=R_x$).
The group characters are $\chi_\pm(I)=1$ and $\chi_\pm(R_x)=\pm 1$,
$S_\gamma$ is the action and $A_\gamma$ is obtained from
the stability of the orbit.
Real orbits cannot cross from one well to the other and
necessarily correspond to $g_\gamma=I$. Their contribution vanishes
when we calculate the difference, a fact which is true to all orders
in $\hbar$. $f(E)$ is then approximated by orbits
corresponding to $g_\gamma=R_x$, which are necessarily complex, 
and from which we can extract the exponentially small splittings.

After substituting the contributions
of complex orbits into the trace formula and integrating, 
we arrive at the following sum \cite{us},
\begin{equation}\label{TheMan} f(E) \approx 
		{2\over\pi}\;\Im \sum_\gamma \;
			\beta_\gamma\;
      { {e^{iS_\gamma/\hbar}} \over {\sqrt{-\det(M_\gamma-I)}} }.
\end{equation}
Here $S_\gamma$ and $M_\gamma$ are respectively the complex action
and monodromy matrix of $\gamma$. ($M_\gamma$ includes, if necessary, a factor 
representing linearisation in the surface of section of the symmetry 
operation $g_\gamma$.) We use a complex square 
root in the denominator and the ambiguity in sign is
determined by following the evolution of the square root in the complex 
plane --- this mirrors the computation of Maslov indices
for real orbits \cite{crlr}. The factor $2$ reflects a degeneracy in the
direction of tunnelling and the dimensionless factor $\beta_\gamma$, 
explained below, is $1$ for most orbits.

We explore this approximation for the two-dimensional potential
\be \label {smokepot}
V(x,y) = (x^2-1)^4 + x^2y^2.
\ee
When $E<1$, the classical motion is confined to one of two symmetric
wells, leading to splittings. 
There is a real periodic orbit confined to the $x$-axis which has a
bifurcation at $E_c=0.236$ such that it is elliptic for $E<E_c$ and
inverse-hyperbolic for $E>E_c$ (in which energy range phase space is
predominantly chaotic). Tunnelling in this system is dominated by a
particular set of complex orbits that are similarly confined to the
$x$-axis. Their contributions are found by first considering the
analogous one-dimensional orbits in the potential $V(x,0) =
(x^2-1)^4$. They are then dressed with the complex monodromy
matrix obtained by embedding them in the other dimension. This is
useful because a very careful analysis of such orbits is possible;
the results of which can later be extended to more general orbits.

The simplest complex orbit starts on the $x$-axis with negative
kinetic energy and evolves after an imaginary time
$i\tau$ to the symmetry-related point on the other side. It has
real position and imaginary momentum, giving an
imaginary action $S=iK$. The amplitude of this orbit is 
reduced by a factor of $2$ on reflection from the inside of the 
barrier. This anomalous reflection coefficient derives from
a calculation in which a Stokes constant is calculated on a Stokes
line \cite{us,Dingle}, and can be understood in simple terms by 
comparison with an exact calculation for the inverted harmonic 
oscillator \cite{us}. As a result, the orbit contributes with a 
factor $\beta_\gamma=1/2$. The monodromy matrix  
can be found by inverting the potential. The tunnelling orbit 
then transforms to a real, unstable orbit running along the
ridge $y=0$, whose monodromy matrix $M$ is readily calculated.
Reverting to the upright potential, the monodromy matrix
$W$ is obtained from $M$ by multiplying the off-diagonal 
elements by $\pm i$. This leaves eigenvalues unchanged. After 
including a phase factor from reflection inside the barrier, we 
obtain
\begin{equation}\label{Mean} f_0(E) = 
		{1\over\pi}\;
      { {e^{-K/\hbar}} \over {\sqrt{-\det(W-I)}} }.
\end{equation}
A final complication derives from the fact that we will
consider seperately states with even and odd $y$-parity.
The  orbit lies on the corresponding symmetry axis and 
contributes differently to the two parities as a result.
The amplitude must then be decomposed according to
the prescription in \cite{bent}.

For comparison with the theory, we also found
the quantum eigenvalues corresponding to the potential
(\ref{smokepot}) numerically. We used 100 harmonic oscillator basis
states in both the $x$ and $y$ directions and diagonalised the resulting
Hamiltonian. We worked to quadruple precision which allowed us to
calculate splittings as small as about $10^{-30}$. All results shown
are for $\hbar=0.01$. We also used the appropriate bases to isolate
the four symmetry classes. States even and odd with respect to $x$ generate
the splittings.
The symmetry with respect to $y$ is additional and we refer
to the corresponding parity classes as even and odd.

\Postscript{7.0}{1.0}{Fig1.eps}{-1.0}{16.5}
{The dots show the quantum splittings multiplied by the
Thomas-Fermi density of states and the solid curves show the results using
$f_0(E)$ of Eq.~(6). ``Even'' and ``odd'' refer to the $y$-parity.}{raw}


In Fig.~1 we show as filled dots the exact splittings $\Delta E_n$
weighted by the Thomas-Fermi density of states $\rho_0(E_n)$
--- these are dimensionless numbers expressing the splittings in units
of the mean level spacing.  The solid curve represents the prediction in
(\ref{Mean}) and agrees well with the global trend of the actual
splittings. There is a deviation near $E=0$ but this is to be expected
because the tunnelling orbit approaches marginal stability there.
Elsewhere the agreement is impressive. In particular,
the theory works for $E<E_c$, in which range the motion is 
mixed and the real one-dimensional orbit stable.
Therefore, the theory predicts the mean behaviour whenever the
tunnelling is dominated by isolated orbits, regardless of the
character of the classical motion in the wells.

In addition to the mean behaviour, there is an oscillatory structure in
the splittings. To explain this,
we consider orbits obtained by attaching, to the basic tunnelling orbit 
above, real periodic orbits in the wells on either side --- always with 
$y=0$. We denote by $S_0$ and $T_0$ the action and period of
the real primitive orbit segment. Restricting the sum to orbits which
tunnel only once, we allow any number of iterations of the real orbit in
the first well before tunnelling and in the second well after tunnelling.
Orbits going from left to right with a total action $rS_0+iK$ occur with a
degeneracy $r$. On doing the trace integral in the wells, we get a
contribution to $g_+(E)-g_-(E)$ with an amplitude factor $rT_0/(i\hbar)$.
Summation over these
contributions is sufficient in one-dimensional calculations to identify 
poles in $g_\pm(E)$ and obtain individual splittings \cite{miller,SF6}. 
For a consistent approximation of $f(E)$, however, it is also necessary 
to add a term with amplitude $i\tau/(i\hbar)$ that arises from
integration across the forbidden region --- a contribution which was
not included (or needed) in the one-dimensional calculations. The
integration in energy leading to the staircase functions leaves a
denominator $\partial(rS_0+iK)/\partial E$ which cancels this
combination of periods. Including a factor $2$ to account for the choice
of starting well, the result is (\ref{TheMan}) with $\beta_\gamma=1$.
This argument can be made precise by considering the problem of
an infinite square well with a finite square barrier in the middle
\cite{us}. The equivalent calculations are exact and free
from the ambiguities inherent in comparing exponentially small quantities
in WKB calculations.

The monodromy matrix for an orbit with $r$ real handles is conjugate to
$M_r = W M_0^r$, where $M_0$ is the real monodromy matrix of a primitive
real orbit segment. $M_r$ has complex entries. Collectively, these
orbits contribute
\begin{equation}\label{handles}
	f_\osc(E) = {2\over\pi} \; \Im \; \sum_{r=1}^\infty
           { {e^{(riS_0-K)/\hbar}} \over {\sqrt{-\det(W M_0^r-I)}} }
\end{equation}
to $f(E)$. To determine the branch of the square root, we write
$\sqrt{-\det(M-I)}=\sqrt{\lambda}(1-\lambda^{-2})$, where $M$ is a
surface-of-section matrix with leading eigenvalue $\lambda$,
defined at each point along the orbit. Initially $\lambda=1$
and it grows along the real axis if we start with
the tunnelling segment. With subsequent evolution in real time,
$\lambda$ moves into the complex plane and, when the real segment is
unstable, executes a counterclockwise rotation of approximately $\pi$
with each period $T_0$, this becoming more
exact with every iteration. To this evolution should be added a
counterclockwise rotation of $\pi$ at each turning point.
Asymptotically, each iteration then leads in the unstable case
to a phase factor $(-i)^3$ in the amplitude, reflecting the Maslov 
index $\sigma=3$ of the real orbit \cite{crlr}.

\Postscript{7.0}{1.0}{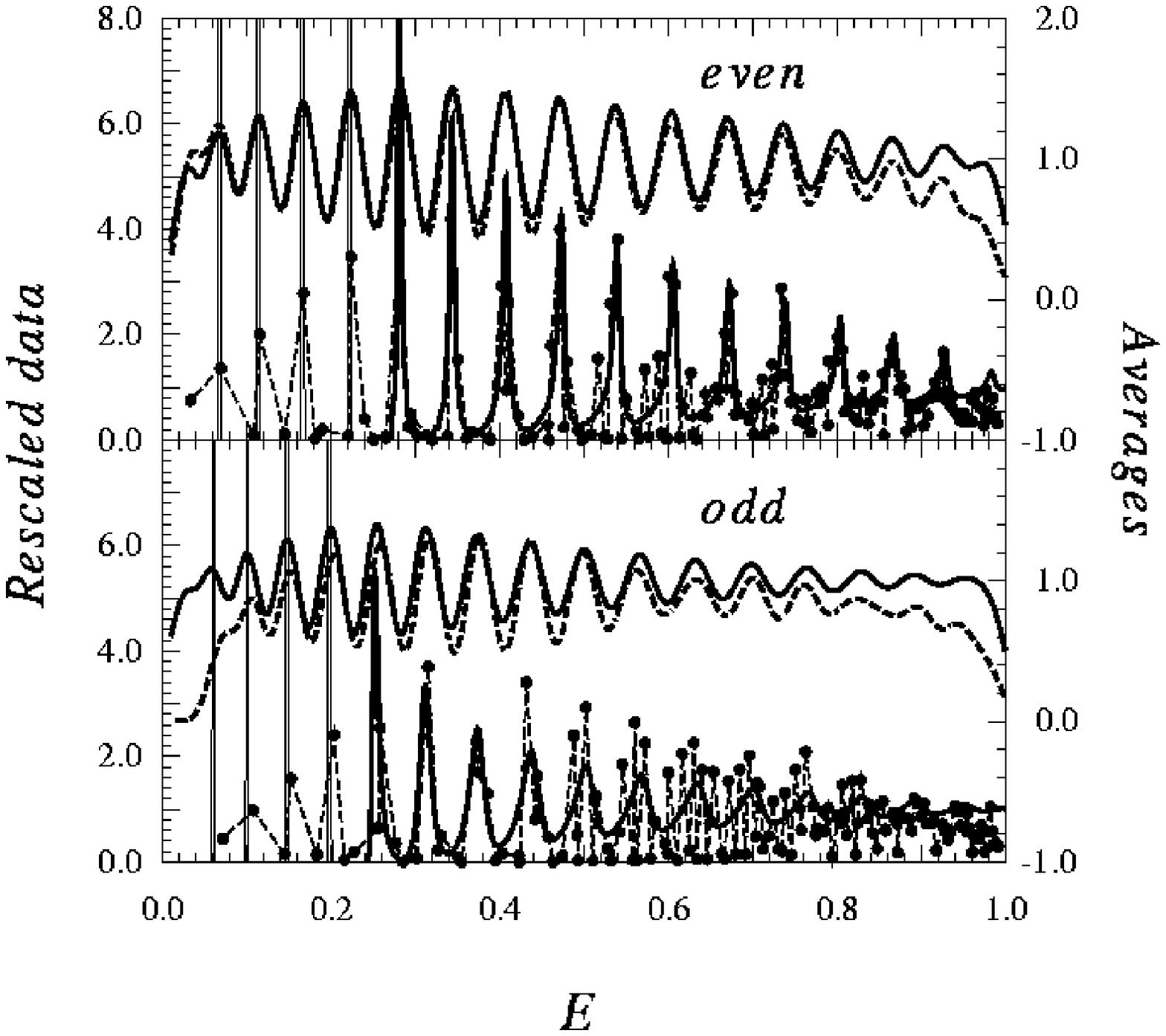}{-1.0}{16.5}
{As in the previous figure but with the mean $f_0(E)$
scaled out. The dots are the quantum data 
with dashed lines connecting them. The lower solid curve is 
$1+f_{\mbox{osc}}(E)/f_0(E)$. The upper dashed and solid curves are
the corresponding results averaged with a Gaussian of width
$w=0.03$. Their scale should be read from the right axis.}{mediumrare}


There are analogies between adding $f_\osc(E)$ to
$f_0(E)$ and adding periodic orbit contributions to the Thomas-Fermi
density of states. In particular the real phase in $f_\osc(E)$
introduces oscillatory structure to the splittings. 
To make a comparison with exact results, it is useful to divide
out the global behaviour apparent in Fig.~1 --- the result should be
a fluctuating function with mean value $1$.
We compare in Fig.~2 the Thomas-Fermi-density-weighted splittings
$\rho_0(E_n)\Delta E_n/f_0(E_n)$ (filled circles) with the function
$1+f_\osc(E)/f_0(E)$ (lower solid curve) obtained by summing over the
first ten repetitions $r$. 

Let us first discuss the energy regime $E<E_c$ where the real
orbit is stable. Here summation over $r$ results 
explicitly in delta functions --- apparent as the first narrow peaks 
in Fig.~2. (In order that the quantum data not be obscured,
side-bands around the peaks have been graphically suppressed.) The 
positions give the 
mean energies of states localised near the orbit in the surrounding island,  
analogous to a quantisation of stable orbits obtained in the usual trace 
formula \cite{poebk}. The peaks have weights of the form 
$(2\hbar A/T_0) e^{-K/\hbar}$ \cite{us}, where $A$ is a function of $W$
and $M_0$. These weights approximate the splittings and are similar
to the one-dimensional case (for which $A=1$). In Table~I are listed
the positions of the peaks and the results of integrating under them.
These predictions are in reasonable agreement with the quantum-mechanical
mean levels and splittings respectively. There are also states which do not 
correspond to peaks in the theoretical curve. We presume that these are
associated with regions of phase space removed from the
orbit. It should be noted that 
in the limit $E\rightarrow 0$, the real orbit suffers an infinite cascade of
bifurcations. Unfortunately, the lowest even state has an energy
near the first such bifurcation,
so the theory cannot be trusted and we do not plot the theoretical
curve in this range.

\begin{table} 
\begin{tabular}{ccccc}
$y$-parity & $E_{\mbox{qm}}$ & $E_{\mbox{sc}}$ & $\Delta E_{\mbox{qm}}$
& $\Delta E_{\mbox{sc}}$\\
\hline
$+$ & 0.0694 & 0.068 & 0.585 $\times 10^{-22}$ & 0.54 $\times 10^{-22}$\\
$+$ & 0.1158 & 0.115 & 0.237 $\times 10^{-20}$ & 0.23 $\times 10^{-20}$\\
$+$ & 0.1677 & 0.167 & 0.883 $\times 10^{-19}$ & 0.91 $\times 10^{-19}$\\
$+$ & 0.2237 & 0.222 & 0.304 $\times 10^{-17}$ & 0.27 $\times 10^{-17}$\\
$-$ & 0.0718 & 0.061 & 0.666 $\times 10^{-23}$ & 0.43 $\times 10^{-23}$\\
$-$ & 0.1076 & 0.101 & 0.212 $\times 10^{-21}$ & 0.19 $\times 10^{-21}$\\
$-$ & 0.1531 & 0.148 & 0.755 $\times 10^{-20}$ & 0.72 $\times 10^{-20}$\\
$-$ & 0.2038 & 0.199 & 0.275 $\times 10^{-18}$ & 0.22 $\times 10^{-18}$
\end{tabular}
\caption{Exact and semiclassical energies and splittings.}
\label{geomsym}
\end{table}

At energies above $E_c$, the real orbit is unstable and phase space
dominated by chaos. The theory has smooth oscillations and there are
no longer individual states associated with these orbits. However, the
theory does reproduce with some detail a marked periodicity in the
splittings. Such oscillations have recently been observed experimentally
by Wilkinson {\it et al} \cite{scars} in quantum wells and interpreted by 
them as corresponding to enhanced tunnelling in states ``scarred'' by 
a real orbit. In our formalism, we obtain an explicit quantitative
prediction, but for averaged tunnelling properties and not (in the
chaotic regime) for individual states. It is hoped that inclusion of
more orbits will yield individual splittings. For a quantitative comparison, 
we compare in Fig.~2  the results of averaging quantum-mechanical
and semiclassical rescaled quantities with a Gaussian of width $w=0.03$.
As discussed before, the theory over-estimates the splittings near $E=1$;
this can in principle be corrected by existing theories for bifurcations
\cite{uniform}.
Also, the cascade of bifurcations at small energies means that the theory is 
not to be trusted in this range either. Other than these  effects, 
it is apparent that we are
correct with regard to the period, amplitude and phase of the
oscillations and therefore have a rather detailed understanding of 
the splittings. Notice also that, as far as averaged
quantities are concerned, there is no qualitative difference between the
regimes in which the orbit is stable or unstable.

The problem studied here made use of a class of orbits which are
essentially one-dimensional --- their two-dimensional character being
carried exclusively through the monodromy matrix. This is an
artifact of the reflection symmetry about the
$y$-axis and it will be interesting to follow these orbits as the
symmetry is broken, either by adding another term to the potential or
by adding a magnetic field. This would be the first step in
understanding how to combine the contributions of many complex orbits
so as to get individual splittings in the chaotic case.
Another necessary ingredient is some form of symbolic dynamics of the complex
orbits so that we have a systematic way of searching for orbits; without
this, it is an extremely difficult task to find long orbits.
A distinct problem is to determine whether the statistics of the splittings
conform to any universal distributions.

The form of analysis developed here may help in understanding
mesoscopic systems such as Coulomb blockade peaks in which 
electrons tunnel into a quantum dot, and on which to date only statistical 
analyses have been brought to bear \cite{cbp}. For this purpose, we note
that a very natural extension of the theory developed here would allow
us to find the width-weighted density of resonances for unbounded
problems. The formalism might also be used to investigate leakage
from or between billiards with small holes. If the holes are small enough,
the contribution of leaking orbits will be suppressed diffractively
so that the widths or splittings will decay algebraically with 
wavelength, as opposed to the exponential decay found in tunnelling.

We would like to thank E. Bogomolny, O. Bohigas, P. Leboeuf, 
A. Mouchet A. Voros for useful discussions.
N.D.W. acknowledges support from the Natural Sciences and 
Engineering Research Council of Canada.

\end{document}